\documentclass[conference]{IEEEtran}
\IEEEoverridecommandlockouts
\usepackage{cite}
\usepackage{amsmath,amssymb,amsfonts}
\usepackage{graphicx}
\usepackage{textcomp}
\usepackage{algorithm}
\usepackage{algpseudocode}
\usepackage{xcolor}
\usepackage{bm}
\def\BibTeX{{\rm B\kern-.05em{\sc i\kern-.025em b}\kern-.08em
    T\kern-.1667em\lower.7ex\hbox{E}\kern-.125emX}}
\begin{document}

\title{Random Channel Ablation for Robust Hand Gesture Classification with Multimodal Biosignals\\
\thanks{$^*$This work was performed while K.B. was an intern at MERL.}
}

\author{
    \IEEEauthorblockN{Keshav Bimbraw\IEEEauthorrefmark{1}\IEEEauthorrefmark{2}, Jing Liu\IEEEauthorrefmark{2}, Ye Wang\IEEEauthorrefmark{2}, and Toshiaki Koike-Akino\IEEEauthorrefmark{2}}
    \IEEEauthorblockA{\IEEEauthorrefmark{1}Worcester Polytechnic Institute, Worcester, MA 01605, USA\\
     \IEEEauthorrefmark{2}Mitsubishi Electric Research Laboratories (MERL),
Cambridge, MA 02139, USA\\
    kbimbraw@wpi.edu, \{jiliu, yewang, koike\}@merl.com}
}

\maketitle

\begin{abstract}
Biosignal-based hand gesture classification is an important component of effective human-machine interaction. For multimodal biosignal sensing, the modalities often face data loss due to missing channels in the data which can adversely affect the gesture classification performance. To make the classifiers robust to missing channels in the data, this paper proposes using Random Channel Ablation (RChA) during the training process. Ultrasound and force myography (FMG) data were acquired from the forearm for 12 hand gestures over 2 subjects. The resulting multimodal data had 16 total channels, 8 for each modality. The proposed method was applied to convolutional neural network architecture, and compared with baseline, imputation, and oracle methods. Using 5-fold cross-validation for the two subjects, on average, 12.2\% and 24.5\% improvement was observed for gesture classification with up to 4 and 8 missing channels respectively compared to the baseline. Notably, the proposed method is also robust to an increase in the number of missing channels compared to other methods. These results show the efficacy of using random channel ablation to improve classifier robustness for multimodal and multi-channel biosignal-based hand gesture classification.
\end{abstract}


\section{Introduction}
Hand gesture recognition using biosignals has been a focus of research for designing pipelines for effective human-machine interfaces. For this, several biosignal modalities have been explored such as surface electromyography (sEMG)~\cite{9751758}, ultrasound~\cite{9812287}, force myography (FMG)~\cite{7908995}, photoplethysmography (PPG)~\cite{10036409}, mechanomyography~\cite{9146882}, inertial measurement unit (IMU)~\cite{9283231}, and electrical impedance tomography (EIT)~\cite{9627682}. Forearm ultrasound can help with visualization of the musculature which can be used to estimate hand gestures~\cite{10331075}. Recent research has shown that it can be used to estimate different hand configurations~\cite{9176483}, finer finger movements~\cite{9812287} and forces~\cite{10306652}. With FMG, piezoelectric sensors around the forearm are used to acquire pressure information corresponding to muscle contraction, from which hand gestures can be inferred. Ha et al. showed an average classification accuracy of over 80\% for estimating hand gestures and prosthetic hand control~\cite{ha2019performance}. Multimodal biosignal-based gesture classification has been explored primarily with sEMG~\cite{gao2021hand}, in addition to modalities like ultrasound~\cite{wei2022multimodal}, FMG~\cite{jiang2020novel} among others. Multimodal data based systems have been shown to improve the gesture classification performance compared to using just one modality \cite{abavisani2019improving}.
We explore using ultrasound and FMG in a multimodal fashion for gesture recognition in this paper. 

With the advancements in machine learning, there is a growing interest in using biosignals for hand gesture classification not just in academia but also in industry~\cite{tchantchane2023review}. The robustness of the machine learning models is of prime importance for effective hand gesture classification. Though combining multiple modalities and multiple channels can often lead to more accurate predictions, it also increases the fragility of the system, as in practice there are often missing channel and missing modality issues due to the unreliability of the sensors and communications, as well as partial occlusions and other disruptions. A common practice is to impute the missing values with zeros or the mean~\cite{DONDERS20061087}. However, such ad-hoc approaches have no robustness guarantee and still suffer from significant performance degradation when the number of missing channels increases. In this paper, we aim to improve the robustness of multimodal and multi-channel signal processing applied to biosignal based hand gesture classification. We leverage the Randomized Ablation technique~\cite{levine2020robustness} from the Adversarial Machine Learning literature, which was proposed to defend against sparse adversarial attacks, where some pixels of an RGB image are adversarially perturbed. 

In this paper, we generalize Randomized Ablation to multi-channel and multimodal settings, which we call Random Channel Ablation (RChA). We apply this to biosignal processing, with a focus on robust hand gesture prediction in the presence of missing channels. We use multimodal ultrasound and FMG data for the classification of 12 hand gestures with a convolutional neural network (CNN). Section II describes the data acquisition, model, gesture classification workflow, and random channel ablation technique. Section III describes the experimental setup for data acquisition, processing, experimental design, and evaluation metrics. Finally, the results and discussion sections describe the obtained results and major takeaways in addition to future research directions stemming from this work.

\begin{figure}[t]
\centerline{\includegraphics[width=200pt]{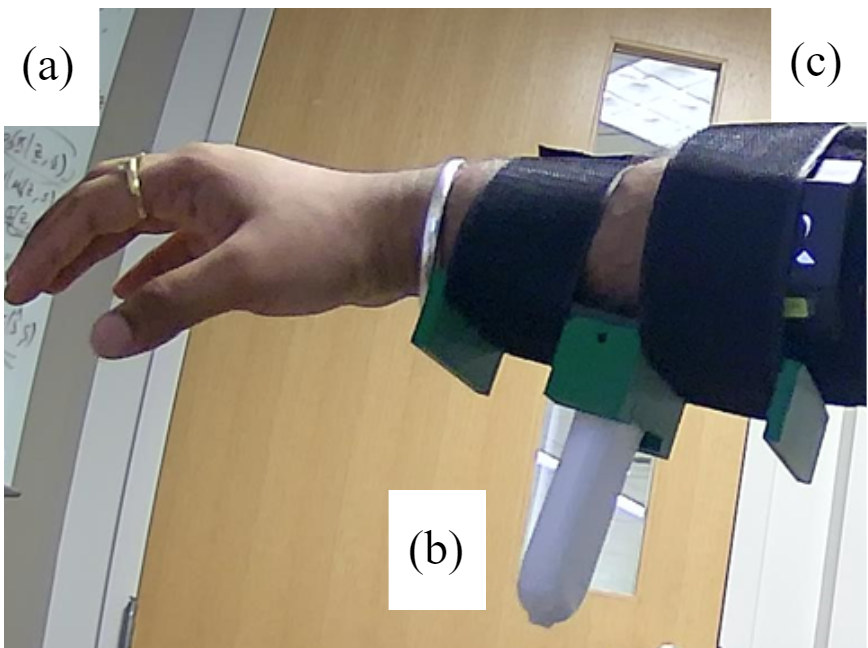}}
\caption{Data acquisition: (a) Hand gesture; (b) Linear ultrasound probe strapped to the forearm using a custom-designed attachment; (c) BioX FMG armband strapped to the forearm.}
\label{data_acq}
\end{figure}

\section{Methods and Experimental Design}

Ultrasound and FMG data were simultaneously acquired from 2 subjects.
The data acquisition procedure and the 12 hand gestures were explained to the subjects. The study was approved by the institutional research ethics committee at Mitsubishi Electric Research Laboratories (IRB reference number 23001), and written informed consent was given by the subjects before the data acquisition. 

\subsection{Data Acquisition}
The ultrasound data was acquired using a Sonostar 4L linear palm Doppler ultrasound probe. A custom-designed 3D-printed wearable was strapped onto the subject's forearm. The data from the probe was streamed to a Windows system over Wi-Fi, and screenshots of the ultrasound images were captured using a custom Python script. The 4L linear probe has 80 channels of ultrasound data, and the post-processed beamformed B-mode data is obtained, from which $350 \times 350$-pixel image data is acquired.
The FMG data was acquired using a BioX AAL-Band 2.0. The data were streamed to the Windows system over Bluetooth, and saved along with the ultrasound frames using the Python script to ensure that the ultrasound and FMG data were synchronized. The FMG band gives 8 channels of FMG data. 
Fig.~\ref{data_acq} shows the data acquisition from the ultrasound and FMG.

The data were acquired for 12 hand gestures selected based on activities of daily living, as described in detail in~\cite{10020174}.
These hand gestures consist of 4 finger flexions, 4 pinch configurations, hand wolf, fist, hook, and open hand gestures. 100 frames of data were acquired while the subject had a fixed hand gesture  Per subject, 24{,}000 frames of data were used for the study which resulted from 20 sessions of data acquisition (20 sessions, 12 hand gestures, 100 frames per gesture: 24{,}000 frames). The average frame rate of data acquisition was 18 Hz. The experiments were run on a desktop with an Intel i7-13700K CPU, 128GB RAM, and an NVIDIA GeForce RTX 4090 GPU.

\begin{figure}[t]
\centerline{\includegraphics[width=230pt]{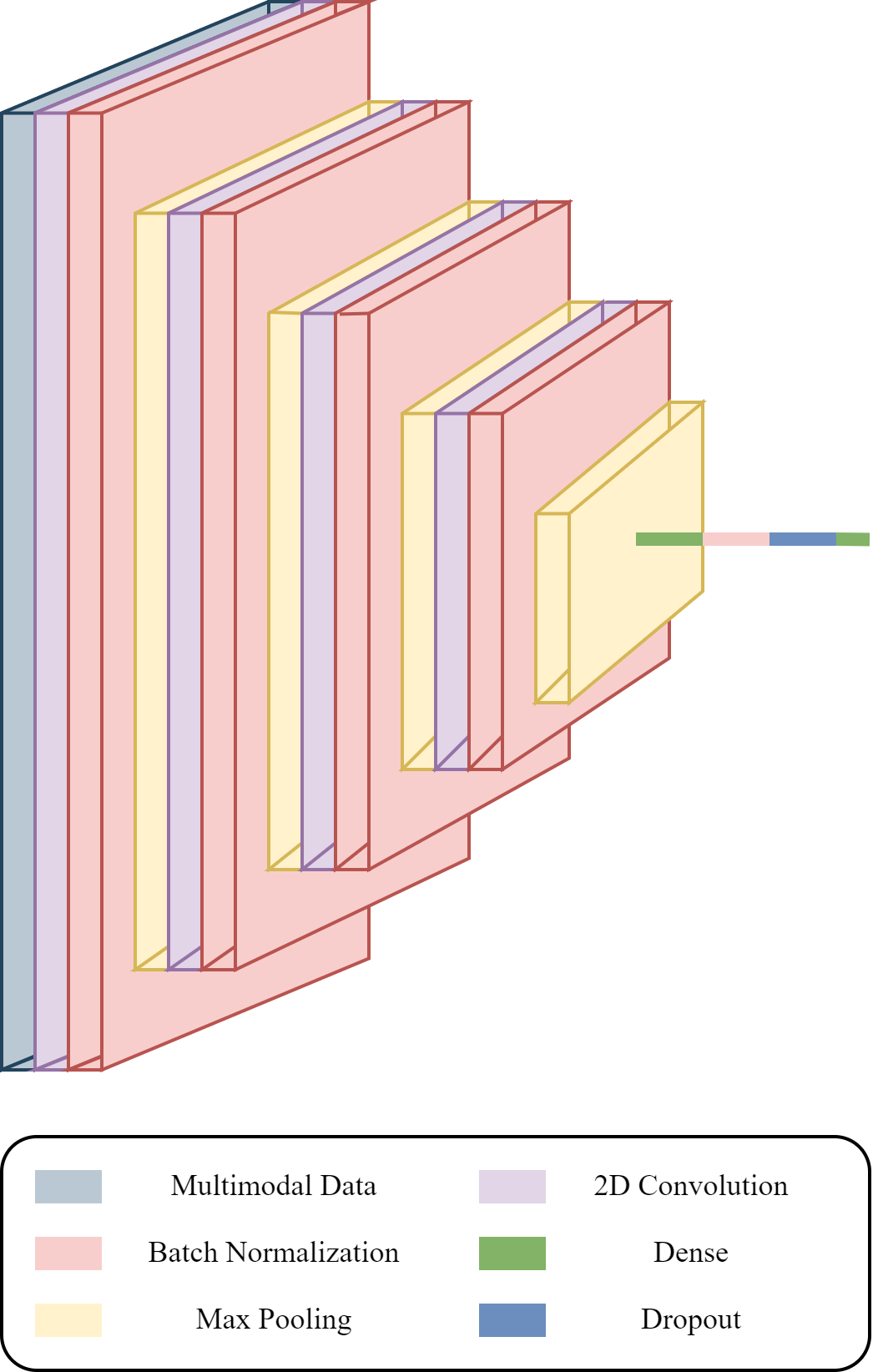}}
\caption{The model architecture to classify hand gestures from multimodal data.}
\label{cls}
\end{figure}

\subsection{Model Architecture, Parameters and Metrics}
A CNN was used in this paper designed using PyTorch based on a network used to classify hand gestures from ultrasound images from the forearm~\cite{10020174, 9812287}. The network is tailored for image compression, and its architecture is designed to capture hierarchical features through convolutional and pooling operations. The network has 4 cascaded convolution layers followed by batch normalization and pooling layers compared to the referenced architecture which has 5 cascaded convolution layers and a different number of parameters per layer because of different input dimensions. The 2-dimensional convolution layers have 16 channels as the output channels, and a kernel size of (3, 3), with a stride of 1. These are followed by a fully connected linear layer which flattens out the output from the fourth max-pooling layer. After another batch normalization, a dropout operation is applied with dropout probability of 0.5. This is followed by a final fully connected layer with a softmax activation which leads to the probabilities of the 12 hand gestures based on the input data. Each convolution layer has a rectified linear unit (ReLU) activation. The first fully connected (dense) layer has a ReLU activation, while the output layer has a softmax activation. The network is visualized in Fig.~\ref{cls}.

\begin{figure*}[t]
\centerline{
  \includegraphics[width=500pt]{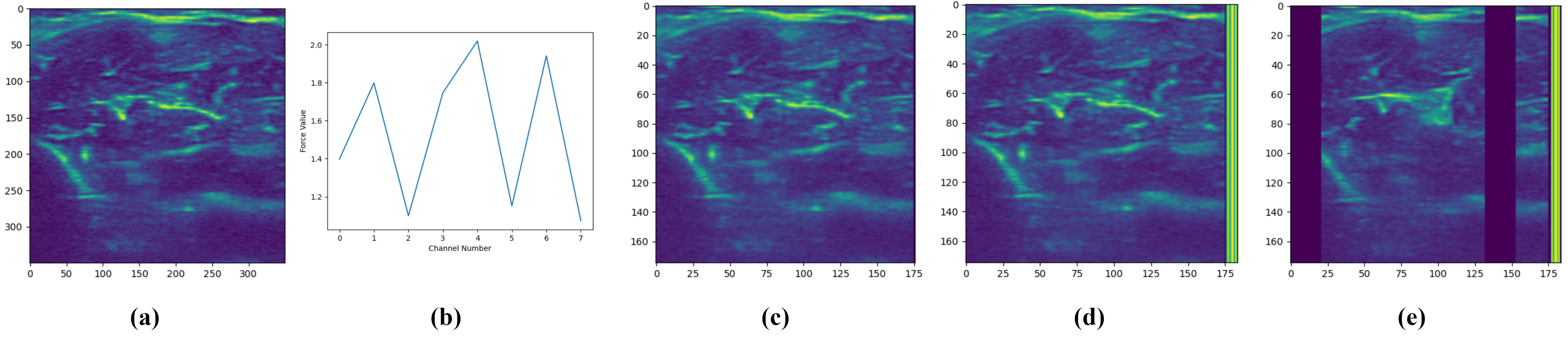}}
  \caption{Data pre-processing and ablation. 
  (a) The $350 \times 350$-pixel forearm ultrasound image was acquired using a linear ultrasound probe. (b) 8-channel FMG data was simultaneously acquired. (c) The ultrasound image was downsized by a factor of 2. The new image dimensions were $175 \times 176$. An additional column of zeros was added to make the column width divisible by 8. (d) Ultrasound and FMG data were independently normalized and then appended per frame. The FMG data was expanded to span the rows of the ultrasound image. The dimensions of each sample were $175 \times 184$ with the last 8 columns belonging to FMG and the remaining to the downsized ultrasound image. (e) The output after random channel ablation with channels 0, 6, 8, and 15 ablated. First, the number of channels to be ablated was chosen randomly which turned out to be 4. Then, 4 channels were randomly chosen and ablated from the 16 channels.}
\label{preprocessing}
\end{figure*}

Cross Entropy Loss was used for evaluating the loss and the Adam optimizer was used for optimizing the network. A learning rate of 0.0001 and a batch size of 10 were used. The evaluation was done for three different random seed values: 0, 1, and 11, and the results were averaged over these seeds. Classification accuracy percentage (\(Acc\)) was used as the accuracy metric, which is the fraction of correct classifications over the total number of classifications multiplied by 100. 

5-fold cross-validation was used to assess the robustness and generalization performance of our method. The dataset was partitioned into five subsets, training the model on four of these subsets, and evaluating its performance on the remaining subset. This meant, that per subject and fold, the training data had 19{,}200 samples and the testing data had 4{,}800 samples. This process was repeated five times, each time using a different subset for evaluation. Throughout training and evaluation, careful attention was given to maintaining temporal independence and preventing any overlap between the train and test sets. The results were averaged to provide a more reliable estimate of the model's performance and reduce the impact of variability in a single train-test split. 

\subsection{Data Preprocessing}
Fig.~\ref{preprocessing} shows the data pre-processing for the ultrasound and FMG data. The $350 \times 350$-pixel ultrasound data and $1 \times 8$ FMG data are obtained per frame. The ultrasound image is downsized by a factor of 2. Let $I$ be the original image with dimensions $d \times d$. The downsized image, $I_{\text{downsized}}$, with dimensions $\frac{d}{2} \times \frac{d}{2}$, is expressed as follows:
\begin{equation}
\label{downsized}
    I_{\text{downsized}}(i, j) = I(2i, 2j).
\end{equation}

Here, $i$ and $j$ are pixel indices in the downsized image, and $2i$ and $2j$ are the corresponding pixel indices in the original image. The downsized image is shown in Fig.~\ref{preprocessing}(c). The dimension of the downsized ultrasound image is $175 \times 175$ pixels. A zeros column was added to the downsized image to make the columns perfectly divisible by 8, with the updated dimensions being $175 \times 176$ pixels. 

To have consistency in the FMG and ultrasound data, the data from both modalities were independently normalized. Following this, the FMG data was expanded to have the same number of rows as the ultrasound data and then appended to the ultrasound data as shown in Fig.~\ref{preprocessing}(d). The dimension of this image is $175 \times 184$ pixels with the last 8 columns representing data obtained from FMG and the remaining 176 columns belonging to the ultrasound image.

\subsection{Robust Training for Hand Gesture Classification}
It has been previously shown that randomly ablating image pixels can help make a classifier robust to sparse adversarial perturbations that only corrupt very few number of image pixels~\cite{levine2020robustness}. This is because the fraction of corrupted pixels is very small and the vast majority of the pixels are benign. There is a high probability that the randomly retained (non-ablated) pixels are not corrupted. Further, it is possible to recognize images with partial observation (i.e., retained pixels). If a classifier is trained deliberately with randomly ablated information (i.e., partial observation), using the Randomized Ablation strategy can be certifiably robust to such sparse adversarial perturbations. In our work, instead of an adversary, we only consider randomly missing channels, and our robust classifier is trained on randomly observed/retained channels.

Algorithm 1 describes the robust training procedure of the proposed Random Channel Ablation method. The pre-processed data is first combined for both modalities. The maximum number of ablated channels $K$, the batch size $M$ for training, and the learning rate $\eta$ are specified. Each batch $X_\mathrm{batch}, \bm{y}_\mathrm{batch}$ in $X_\mathrm{train}, \bm{y}_\mathrm{train}$ is then used for training the neural network and its parameters $\bm \theta$ are then updated. The parameters are updated based on the gradient $\nabla_{\bm \theta}$ of the loss function w.r.t. the model parameters $\bm \theta$. The loss function $\ell(f_{\bm \theta}(X_\mathrm{ablate}),\bm{y}_\mathrm{batch})$ is based on the loss obtained from the predictions $f_{\bm \theta}(X_\mathrm{ablate})$ and the true labels $\bm{y}_{batch}$. 

\begin{algorithm}[t]
    \begin{algorithmic}[1]
    \caption{Robust training via Random Channel Ablation of $T$ epochs. The neural network $f$ is parameterized by \textbf{$\bm{\theta}$}. }
        \State Input $X_\mathrm{train} = $ append($X_\mathrm{us}, X_\mathrm{fmg}$), $\bm{y}_\mathrm{train}$, max \# of ablated channels $K$, batch size $M$, learning rate $\eta$
        \For{$t=1,\ldots,T$}
            \For{\{$X_\mathrm{batch}, \bm{y}_\mathrm{batch}$\} in \{$X_\mathrm{train}, \bm{y}_\mathrm{train}$\}}
 
                \State $X_\mathrm{ablate}\gets$ ablate($X_\mathrm{batch}, K$)
                \State $\bm \theta = \bm \theta - \eta \nabla_{\bm \theta} \ell(f_{\bm \theta}(X_\mathrm{ablate}),\bm{y}_\mathrm{batch})$
                \EndFor
        \EndFor
        \State \textbf{return} neural network parameters $\bm \theta$
\\
        \State \textbf{Subroutine} {$X_\mathrm{ablate}\gets$ ablate($X_\mathrm{batch}, K$)} :\\
        Initialize $X_\mathrm{ablate} = X_\mathrm{batch}$
        \State Randomly choose \# of ablated channels $k \in [0,\ldots,K]$
        \If {$k>0$,}       
        randomly choose $k$ channels of $X_\mathrm{ablate}$ and set them to zero \EndIf
        \end{algorithmic}
\label{alg1}
\end{algorithm}
 
Fig.~\ref{preprocessing}(e) shows the multimodal data sample after randomly ablating 4 channels. First, the number of channels to be ablated was set as 4. Then, 4 channels were randomly chosen and ablated from the total 16 channels. For FMG, each column corresponds to one channel, and for ultrasound, each channel is 22-pixel wide column. This was done so that the ablation was consistent across the modalities.

\subsection{Evaluation}
During testing, the index of missing channels is chosen uniformly at random for each sample. For missing channels, the proposed method fills zero to the missing channels, and then applies robust classifier trained in Algorithm 1 for prediction.

The proposed method is compared with the Baseline method, Imputation method, and Oracle, where the classifier of these 3 methods is trained on all channels. The Oracle has all channels available during testing as well, and its performance can be viewed as an upper bound for the missing-channel case. To deal with missing channels at testing, the Baseline method simply fills in with zero values. While for the Imputation method, the missing channel is imputed using the overall sample mean of that channel from the training samples.

\section{Results} 
Upon averaging the hand gesture classification results over the two subjects, the average Oracle case hand gesture classification accuracy was 92.2\%. Table~\ref{tab:my_label1} shows the gesture classification accuracy for the Baseline, Imputation, Proposed, and Oracle methods when the fraction of missing channels is up to 25\%. Since the multimodal data comprises 16 channels, this leads to a maximum of 4 channels being missing (i.e., the number of missing channels of each testing sample is randomly chosen from 1 to 4). On average, our proposed method's classification accuracy was 90.9\%, compared to 88.8\% for the Imputation method and 78.7\% for the Baseline method. 

\begin{table}[t]
\caption{Classification accuracy for different methods when samples are randomly missing up to 25\% channels}
   
    \centering
    \begin{tabular}{ccccc}
    \hline
          & Baseline &Imputation& Proposed & Oracle \\
         \hline
        Subj 1 & 70.0\%&81.9\% & \textbf{86.2}\% & 87.3\% \\
        \hline
        Subj 2 & 87.3\%& \textbf{95.6}\% & {95.5}\% & 97.1\%\\
        \hline
        Average & 78.7\% & 88.8\% & \textbf{90.9}\% & 92.2\%\\
        \hline
    \end{tabular}
     \label{tab:my_label1}
\end{table}

Table~\ref{tab:my_label2} shows the gesture classification accuracy for the Baseline, Imputation, Proposed, and Oracle methods when the fraction of missing channels is up to 50\%. This leads to a maximum of 8 channels being missing. On average, our proposed method's classification accuracy was 89.5\% compared to 83.7\% for the Imputation method and 65.0\% for the Baseline method. 

\begin{table}[t]
\caption{Classification accuracy for different methods when samples are randomly missing up to 50\% channels}

    \centering
    \begin{tabular}{ccccc}
    \hline
         &  Baseline &Imputation& Proposed & Oracle \\
         \hline
        Subj 1  & 56.1\% & 75.1\%  & \textbf{83.8}\% & 87.3\% \\
        \hline
        Subj 2  & 73.9\% &92.2\%& \textbf{95.1}\% & 97.1\%\\
        \hline        
        Average  & 65.0\% &83.7\% & \textbf{89.5}\% & 92.2\%\\
        \hline
    \end{tabular}
     \label{tab:my_label2}
\end{table}

\begin{figure}[t]
\centerline{\includegraphics[width=\linewidth]{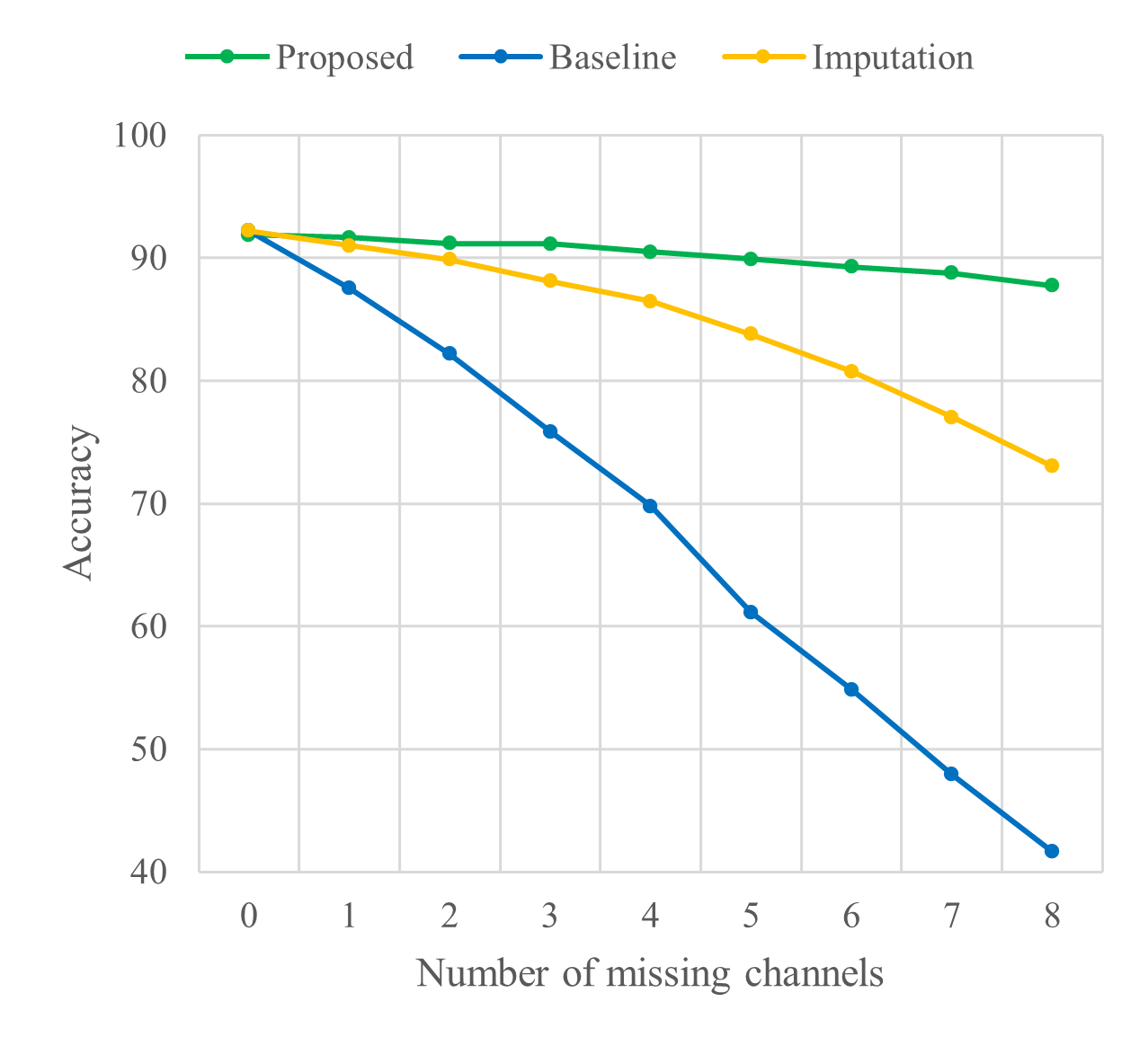}}
\caption{Classification accuracy for a fixed number of channels being missing.}
\label{fig4}
\end{figure}

Fig.~\ref{fig4} shows the results where a fixed number of channels are missing per sample for the different methods. For 4 missing channels, on average, our proposed method's classification accuracy was 90.5\% compared to 86.5\% for the Imputation method and 69.8\% for the Baseline method. For 8 missing channels, on average, our proposed method's classification accuracy was 87.7\% compared to 73.0\% for the Imputation method and 41.6\% for the Baseline method. It is also worth noting that, when there is no missing channel, our robustly trained classifier performs almost the same as the Baseline method (which is also the Oracle method since there are no missing channels). 

\section{Discussion}
In this section, we discuss the effect of random channel ablation, and the effect of the number of missing channels on the classification performance. We also highlight the advantages of our method in making it robust to missing channels and some future research directions stemming from this work.

\subsection{Effect of Random Channel Ablation}

As can be seen in both Tables~\ref{tab:my_label1} and~\ref{tab:my_label2}, the proposed method outperforms the Baseline and the Imputation methods. For a maximum of 4 (25\%) channels being missing, there is an improvement of 12.2\% compared to the Baseline method, and 2.1\% compared to the Imputation method. The classification accuracy is 1.3\% lower than the Oracle case, which is expected since the proposed method only has partial observation of the channels during testing. Similar trends were observed with a maximum of 8 (50\%) channels being missing with an improvement of 24.5\% from the Baseline method and 5.8\% from the Imputation method. The classification accuracy for the proposed method is only 2.7\% less than the Oracle case. This shows that random channel ablation can help make the classifier robust to missing channels, while still maintaining very high utility.

\subsection{Effect of Number of Missing Channels}

With a higher number of missing channels in the test set, it is expected to see a significant drop in classification accuracy, as can be seen for the Baseline method in Fig.~\ref{fig4}. For 4 channels missing, the proposed method's accuracy is 20.7\% better than the Baseline method and 4.0\% better than the Imputation method. For 8 channels missing, the proposed method's accuracy is 46.1\% better than the Baseline method and 14.7\% better than the Imputation method. Overall, the drop in the classification accuracy from 0 channels missing to 8 channels missing is a mere 4.1\% for the proposed method compared to 19.2\% for the Imputation method and 50.6\% for the Baseline method, demonstrating clearly superior performance compared to the other methods.

\subsection{A Universal All-in-One Classifier}

For the proposed random channel ablation method, only one universal classifier needs to be trained, and it can be applied to all situations with up to $K$ out of $n$ channels missing.
In contrast, one could train a specific classifier for the specific missing-channel situation. For example, one could train a specific classifier for the situation where only the second channel is missing. However, this would scale poorly when the number of total channels $n$ and the maximum number of missing channels $K$ is large, since one would have to train $\sum_{k=0}^K\binom{n}{k}$ classifiers in total, which would not be practical.

\subsection{Future Research}

This work focused on showcasing the effectiveness of our random channel ablation technique for multimodal and multi-channel biosignals through ultrasound and FMG modalities. For future work, we aim to diversify the data acquisition by incorporating additional modalities, such as surface electromyography, inertial measurement unit gloves, and cameras, among others. Considering the temporal aspects of the input data can be used to improve the classification of hand gestures, and it can also help leverage some additional modalities in a multimodal setting such as sEMG which has been proven to deliver good gesture classification performance \cite{simao2019emg}. 
For merging data from multimodal ultrasound and FMG information, alternative learning based approaches can be explored such as in \cite{li2024dfn} and \cite{yang2023fusion}. Within the scope of the system described in the paper, these approaches would entail learning a low-dimensional embedding of the ultrasound data which would then be fused with the 1D FMG signal.

Additionally, the current study focused on data acquired from 2 subjects. To ensure the generalizability and applicability of multimodal and multi-channel biosignal based hand gesture recognition using our proposed random channel ablation approach, future work will focus on acquiring data from multiple subjects. The expanded set of subjects will enable an evaluation of the proposed approach for subjects with varying characteristics, making it more versatile.

\section{Conclusion}

In this study, we introduced our random channel ablation technique aimed at enhancing the robustness of deep learning classifiers used for hand gesture recognition when faced with missing channels in multimodal and multi-channel biosignal data. Our approach demonstrated notable resilience to up to 50\% missing channels, surpassing the performance of the Baseline and Imputation methods. We observed minimal degradation in classifier performance compared to the Oracle case. The efficacy of our method unveils promising prospects for its application in diverse multimodal and multi-channel biosignal training scenarios. This technique holds practical implications for real-world data acquisition and deep learning model performance, particularly in scenarios where certain channels may be absent. We illustrated the capability to bolster model robustness which can be utilized for several applications and across several modalities.

\section{Acknowledgments}
The authors are grateful to the Mitsubishi Electric Research Laboratories (MERL) researchers for their input and feedback.

\bibliography{reference}
\bibliographystyle{IEEEtran}

\end{document}